\documentclass{PoS}

\title{Cluster virial expansion for quark and nuclear matter}

\ShortTitle{Cluster virial expansions}

\author{\speaker{D. Blaschke}\\
Institute for Theoretical Physics, University of Wroclaw, 50-204
Wroclaw, Poland \\
Bogoliubov Laboratory for Theoretical Physics, Joint Institute for
Nuclear Research, RU-141980 Dubna, Russia\\
E-mail: \email{blaschke@ift.uni.wroc.pl}}

\abstract{
We employ the $\Phi-$ derivable approach to many particle systems with strong correlations that can lead to the formation of bound states (clusters) of different size. 
We define a generic form of $\Phi-$ functionals that is fully equivalent to a selfconsistent cluster virial expansion up to the second virial coefficient for interactions among the clusters.
As examples we consider nuclei in nuclear matter and hadrons in quark matter, with particular attention to the case of the deuterons in nuclear matter and mesons in quark matter. 
We derive a generalized Beth-Uhlenbeck equation of state, where the quasiparticle virial expansion is
extended to include arbitrary clusters. 
The approach is applicable to nonrelativistic potential models of nuclear matter as well as to relativistic
field theoretic models of quark matter. 
It is particularly suited for a description of cluster formation and dissociation in hot, dense matter.
}

\FullConference{XXII International Baldin Seminar on High Energy Physics Problems,\\
        15-20 September 2014\\
        JINR, Dubna, Russia}

\begin{document}

\section{Introduction}
Let us consider a many-particle system with strong interactions which lead to the formation of bound states (clusters) of different size. One example is nuclear matter, where nucleons are the elementary degrees of
freedom and the clusters such as deuterons, tritons, helions, alphas etc. coexist in a nuclear statistical equilibrium (NSE). Another example is quark matter, where mesons and baryons form a spectrum hadronic    resonances with the specifics that due to quark confinement no free quark states shall get populated at low densities.
The problem we shall consider in this contribution is that the clusters may interact with each other thus forming higher order correlations and their formation itself shall modify the properties of their constituents. 
These aspects have to be dealt with in a consistent way, obeying conservation laws (sum rules) despite the fact that necessarily approximations have to be applied such as truncations of the hierarchy of many-particle Greens functions describing the correlations.

The problem to define conserving approximations to the many-body problem has been solved in general by the so-called $\Phi-$derivable approach \cite{KB} which could be applied to the problem of systems with  composite particles \cite{David2,David3} but in fact has been used in the approximation that all degrees of freedom have no substructure \cite{David4,David5, David6,David1}.
On the other hand, a quantum statistical approach has been developed for the problem of bound state formation (and their dissociation by the Mott effect) in many-particle systems based on a cluster decomposition of the selfenergy \cite{RMS}. 
This approach has been developed further to include both, bound and scattering states of the two-particle spectrum in a dense medium in a consistent way given by a generalization \cite{Zimmermann:1985ji,SRS} of the well-known Beth-Uhlenbeck approach \cite{Beth:1936zz,Beth:1937zz}. 
A next step in this development was the inclusion of larger clusters \cite{Typel:2009sy,Ropke:2014fia} and their mutual interactions to lowest order (second virial coefficient) within a cluster virial expansion \cite{Ropke:2012qv}.

In the present contribution we will show how both approaches can be joined, i.e., how a $\Phi$ functional must be introduced in order to define the cluster virial expansion. 
As an example serves the formation of deuterons in nuclear matter described on the basis of a nonrelativistic Greens function approach employing a separable potential.
Another example is the formation of mesons in quark matter described within the field-theoretic Nambu--Jona-Lasinio model, where the present approach allows to define the contribution of mesons to the quark selfenergy, analoguous to the case of a Dirac fermion coupled to a scalar field \cite{Blaizot:1991kh}, see
also \cite{Kitazawa:2014sga}. 
This is an effect missing up to now in the Beth-Uhlenbeck description 
\cite{Hufner:1994ma,Zhuang:1994dw,Blaschke:2013zaa}
of the thermodynamics of the quark-meson system, see also \cite{Yamazaki:2012ux,Wergieluk:2012gd}.
We give an outlook to the case of higher order clusters in nuclear matter as well as to a consistent description of the system of diquarks, mesons and baryons in quark matter.

\section{Selfconsistent approximation scheme and cluster decomposition 
%$\Phi$ derivable approach
}
\begin{figure}[!htb]
\begin{minipage}{0.35\textwidth}
\includegraphics[width=1\textwidth]{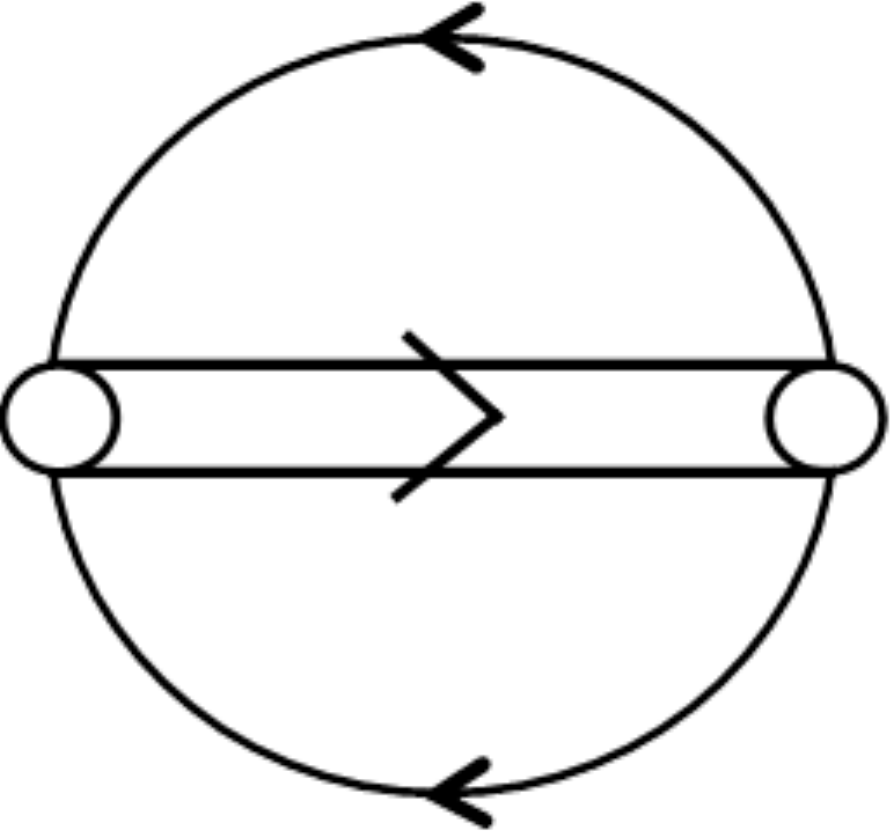}
\end{minipage}\hfill
\begin{minipage}{0.35\textwidth}
\includegraphics[width=1\textwidth]{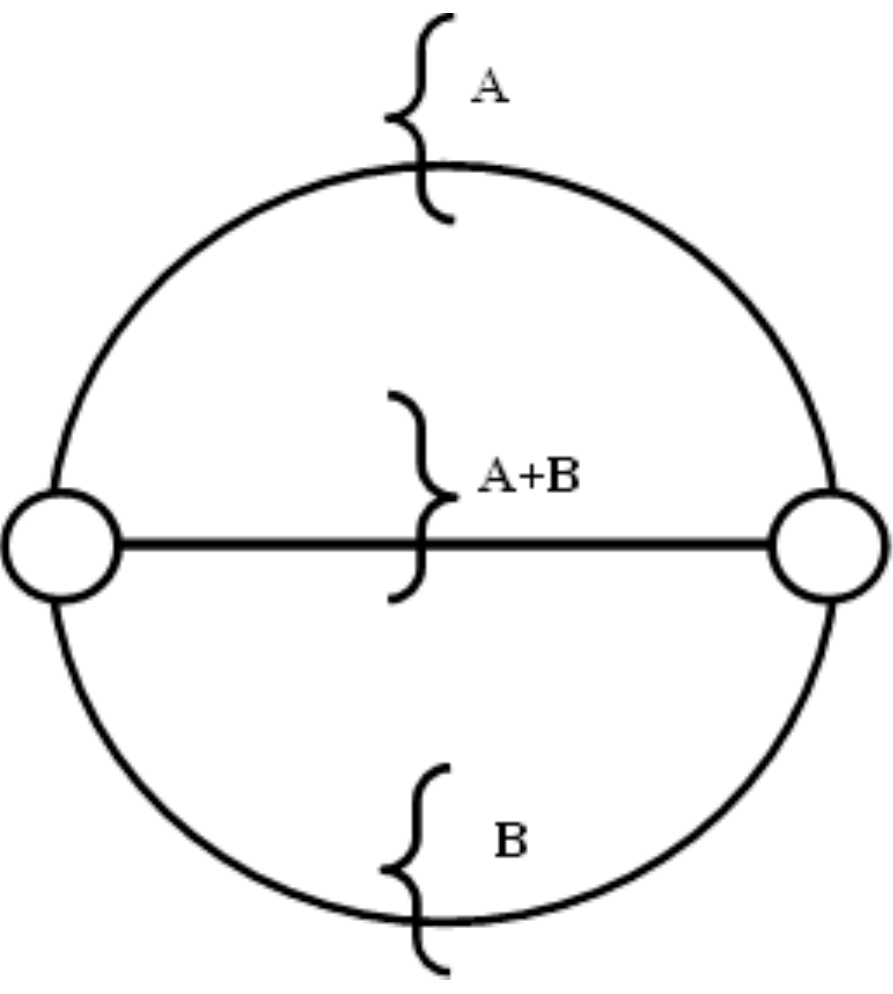}
\end{minipage}
\caption{Left panel: Two-particle irreducible  $\Phi-$ functional describing two-particle correlations (double line with arrow) of elementary fermions (single arrowed lines);
Right panel: The $\Phi-$ functional for the general case of $A-$particle correlations in a many-fermion system.} 
\label{fig:1}
\end{figure}

Within the $\Phi-$ derivable approach \cite{KB} the grand canonical thermodynamic potential 
for a dense fermion system with two-particle correlations is given as
\begin{eqnarray}
\label{2PI}
\Omega = -{\rm Tr}\,\{ \ln (-G_1)\} -{\rm Tr} \{\Sigma_1 G_1\}
+{\rm Tr}\,\{ \ln (-G_2)\} +{\rm Tr} \{\Sigma_2 G_2\} +\Phi[G_1,G_2]~,
\end{eqnarray}
where the full propagators obey the Dyson-Schwinger equations
\begin{eqnarray}
G_1^{-1}(1,z)&=& z - E_1(p_1) - \Sigma_1(1,z);~
G_2^{-1}(12,1'2',z) =z-E_1(p_1)-E_2(p_2)-\Sigma_2(12,1'2',z),
\end{eqnarray}
with selfenergies 
\begin{eqnarray}
\Sigma_1(1,1') &=& \frac{\delta  \Phi}{\delta G_1(1,1')}~;~~
\Sigma_2(12,1'2',z)=\frac{\delta  \Phi}{\delta G_2(12,1'2',z)}~,
\end{eqnarray}
which are defined by the choice for the $\Phi-$ functional, a two-particle irreducible 
set of diagrams such the one in Fig.~\ref{fig:1}.

The functional for the thermodynamic potential (\ref{2PI}) is constructed such that the requirement of its stationarity in thermodynamic equilibrium is equivalent to \cite{David1}
\begin{eqnarray}
\label{n}
 n = - \frac{1}{V} \frac{\partial \Omega}{\partial \mu}
= \frac{1 }{V} \sum_{1}
\int_{-\infty}^\infty\frac{d \omega}{ \pi} f_{1}(\omega) A_1(1,\omega)\,, 
%\nonumber\\
% S_1(1,\omega) &=& {2 {\rm Im}\,\Sigma_1(1,\omega-i0) 
%\over(\omega - E(1)- {\rm Re}\, \Sigma_1(1,\omega))^2 + 
%({\rm Im}\,\Sigma_1(1,\omega-i0))^2 }
\end{eqnarray}
where $A_1(1,\omega)=2 \Im G_1(1,\omega+i\eta)$ is the fermion spectral function and Eq.~(\ref{n})
expresses particle number conservation in a system with volume $V$.  
This approach is straightforwardly generalized to $A-$particle correlations in a many-fermion system
\begin{eqnarray}
\Omega &=& \sum_{A} (-1)^A \left[{\rm Tr} \ln \left(-G_A^{-1}\right) 
+ {\rm Tr} \left(\Sigma_A~G_A \right) \right]+\sum_{A,B}\Phi[G_A,G_B,G_{A+B}]~,
%\nonumber
\\
G_{A}^{-1}&=&G_{A}^{(0)^{-1}} - \Sigma_A~,~~ 
\Sigma_A(1\dots A,1^\prime \dots A^\prime,z_A) = 
\frac{\delta \Phi}{\delta G_A(1\dots A,1^\prime \dots A^\prime,z_A)}~.
%\nonumber
\label{Phi-A}
\end{eqnarray}
The $\Phi-$ functional for this general case is depicted diagrammatically in the right panel of Fig.~\ref{fig:1} 
Herewith we have generalized the notion of the $\Phi$ derivable approach to that of a system where the hierarchy of higher order Green functions is built successively from the tower of all Greens functions starting with the fundamental one $G_1$. The open question is how to define the vertex functions joining the Greens functions. 

\section{Cluster virial expansion within the $\Phi$ derivable approach}

\begin{figure}[!htb]
\begin{minipage}{0.2\textwidth}
\includegraphics[width=1\textwidth]{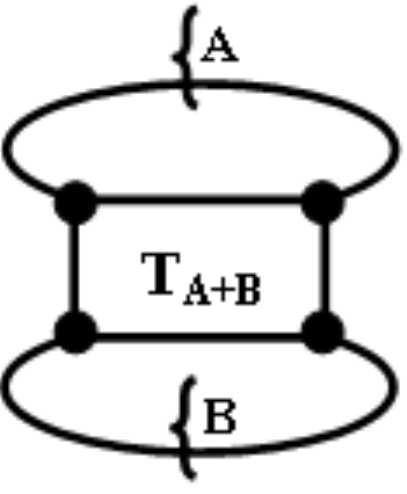}
\end{minipage}
\hfill
\begin{minipage}{0.7\textwidth}
\includegraphics[width=1\textwidth]{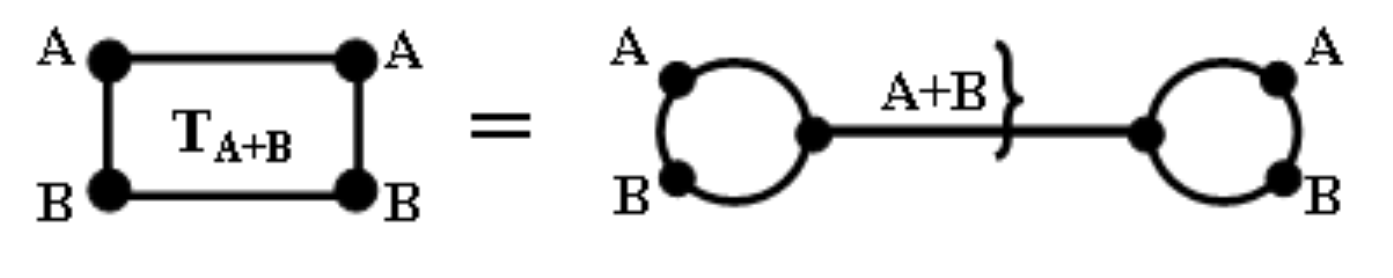}
\end{minipage}
\caption{The $\Phi-$ functional for the cluster virial expansion (left), where the $T_{A+B}-$ matrix for binary collisions in the channel with the partition $A, B$ replaces the higher order Green function $G_{A+B}$ and the corresponding vertex functions according to the scheme given in the right panel.} 
\label{fig:2}
\end{figure}

Having introduced the notion of a cluster expansion of the $\Phi-$ functional we want to suggest a definition which eliminates the unknown vertex functions in favour of the $T_{A+B}-$matrix which describes the nonperturbative binary collisions of $A-$ and $B-$ particle correlations in the channel $A+B$, see 
Fig.~\ref{fig:2}.
In this way we can draw the connection between the cluster virial expansion of Ref.~\cite{Ropke:2012qv}
with the $\Phi-$ derivable approach \cite{KB}.

\subsection{Deuterons in nuclear matter}
The application of this scheme to the simplest case of two-particle correlations in the deuteron channel
in nuclear matter results in the selfenergy \cite{SRS}
\begin{equation}
\label{cluster-SE}
\Sigma(1,z)=\sum_2\int\frac{d\omega}{2\pi} A(2,\omega)\bigg\{ 
f(\omega)V(12,12) -\int\frac{dE}{\pi} \Im T(12,12;E+i\eta) \frac{f(\omega)+g(E)}{E-z-\omega}\bigg\}~,
\end{equation}
where $f(\omega)=[\exp(\omega/T)+1]^{-1}$ is the Fermi function and 
$g(\omega)=[\exp(\omega/T)-1]^{-1}$ the Bose function.
The decomposition (\ref{cluster-SE}) corresponds to a cluster decomposition of the nucleon density 
\begin{equation}
n(\mu,T)=n_{\rm free}(\mu,T) + 2n_{\rm corr}(\mu,T)~,
\end{equation}
where the correlation density contains besides the bound state a scattering state contribution %given by
\begin{equation}
\label{ncont}
n_{\rm sc}=\int \frac{dE}{2\pi} g(E) 2 \sin^2\delta(E) \frac{d\delta(E)}{dE}~.
\end{equation}
The one-particle density of free quasiparticle nucleons is reduced in order to fulfil the baryon number conservation in the presence of deuteron correlations and contains a selfenergy contribution due to 
the deuteron correlations in the medium.
This improvement of the quasiparticle picture due to the correlated medium accounted for by the consistent definition of the selfenergy as a derivative of the $\Phi-$ functional (\ref{Phi-A}) is the reason the continuum 
correlations (\ref{ncont}) are reduced by the factor $2\sin^2 \delta$ as compared to the traditional Beth-Uhlenbeck formula \cite{Beth:1936zz,Beth:1937zz}. For details, see  \cite{Zimmermann:1985ji,SRS}.
With the definition of the $\Phi-$ functional via the T-matrix in Fig.~\ref{fig:2} we were able to show the correspondence between the generalized Beth-Uhlenbeck approach and the $\Phi-$ derivable approach
for the nonrelativistic potential model approach to two-particle correlations in a warm, dense Fermion system \cite{SRS,Ropke:2012qv}.
Now we would like to discuss its application to a relativistic model for Mesons in quark matter.

\subsection{Mesons in quark matter}
\begin{figure}[!htb]
%\begin{minipage}{0.22\textwidth}
\includegraphics[width=0.23\textwidth]{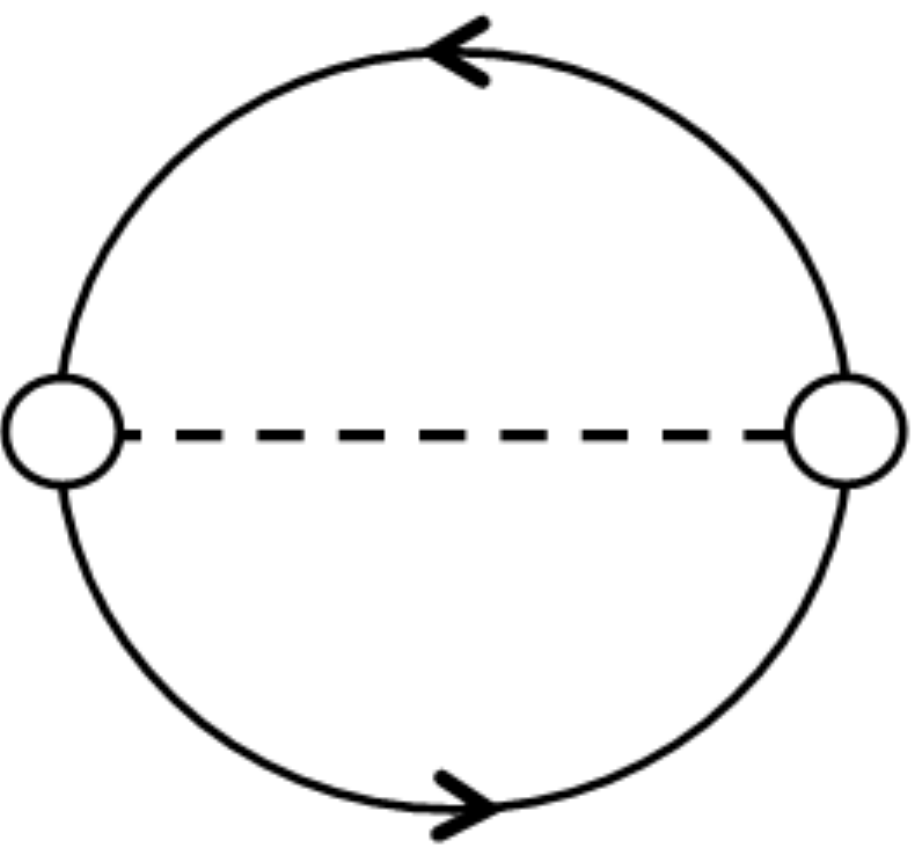}
\hfill
%\end{minipage}
%\begin{minipage}{0.35\textwidth}
\includegraphics[width=0.35\textwidth]{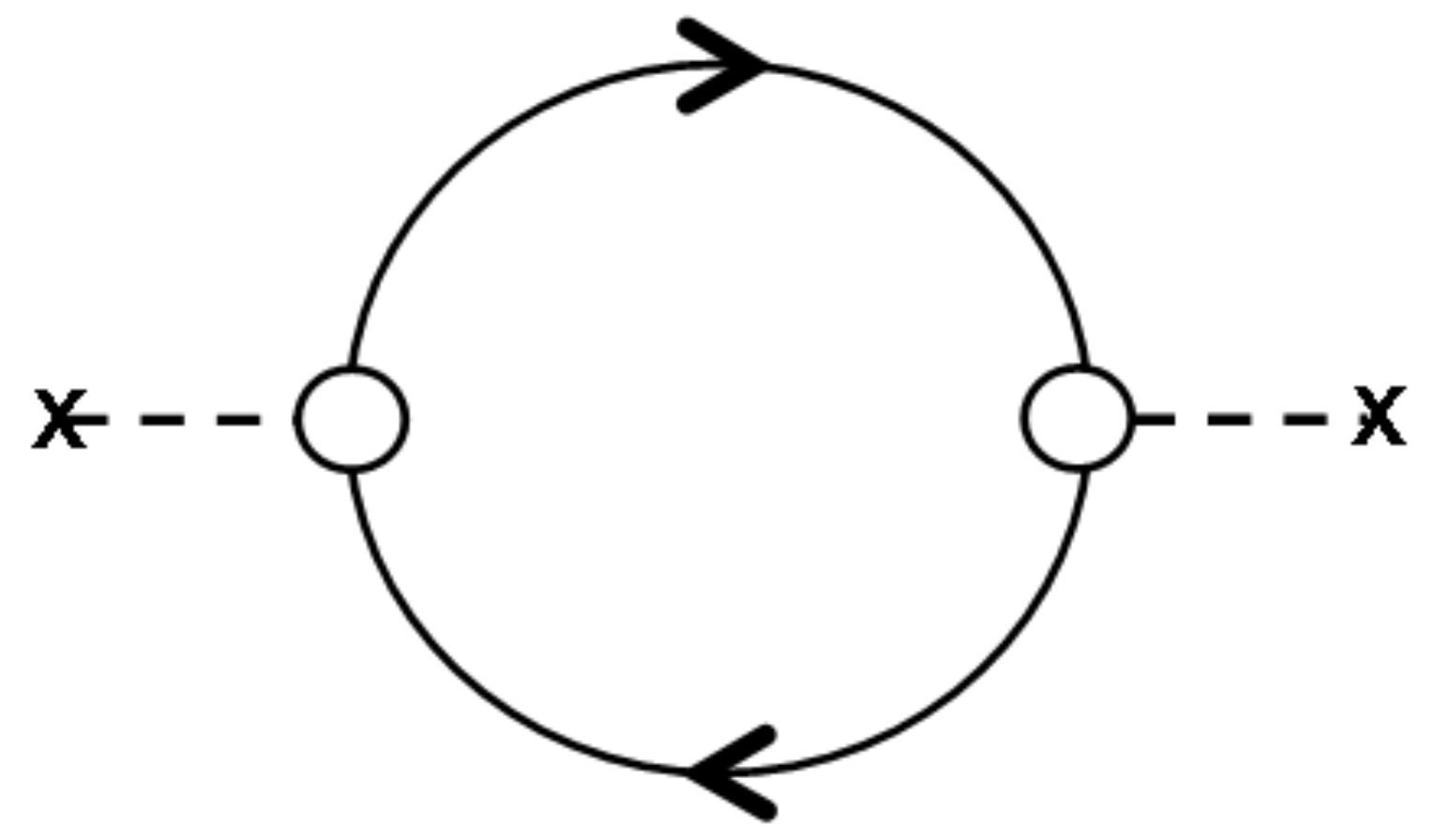}
%\end{minipage}
%\begin{minipage}{0.35\textwidth}
\includegraphics[width=0.35\textwidth]{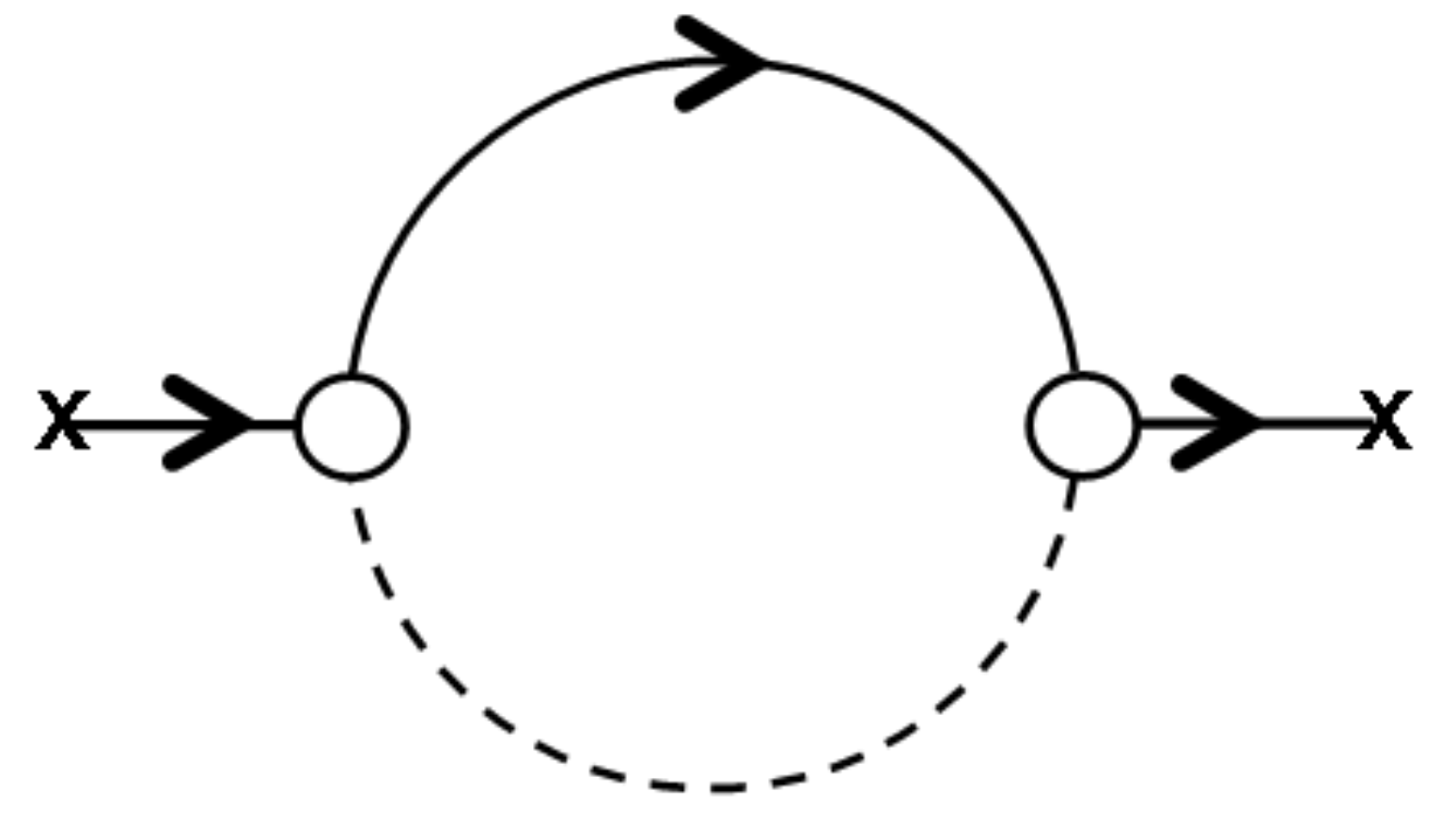}
%\end{minipage}
\caption{The $\Phi-$ functional (left panel) for the case of mesons in quark matter, where the bosonic meson propagator is defined by the dashed line and the fermionic quark propagators are shown by the solid lines with arrows.  The corresponding meson and quark selfenergies are shown in the middle and right panels, respectively.} 
\label{fig:3}
\end{figure}

In order to describe the problem of mesons in quark matter within the $\Phi-$ derivable approach we define 
the $\Phi-$ functional and the corresponding selfenergy in Fig.~\ref{fig:3}.

The meson polarization loop $\Pi_M(q,z)$ in the middle panel of Fig.~\ref{fig:3} enters the definition of the meson T-matrix (often called propagator) 
\begin{equation}
T^{-1}_M(q,\omega+i\eta) = G_S^{-1} - \Pi_M(q,\omega+i\eta)
=|T_M(q,\omega)|^{-1}{\rm e}^{- i \delta_M(q,\omega)}~,
\end{equation}
which in the polar representation introduces a phase shift 
$\delta_M(q,\omega)=\arctan (\Im T_M/\Re T_M)$, that results in a generalized Beth-Uhlenbeck equation of state for the thermodynamics of the consistently coupled quark-meson system \cite{Blaschke:2013zaa}.
\begin{equation}
\Omega = \Omega_{\rm MF} + \Omega_M ~, 
\end{equation}
where the selfconsistent quark meanfield contribution is
\begin{equation}
\Omega_{\rm MF} =
\frac{\sigma^2_{\rm MF}}{4G_S} - 2N_c N_f \int\frac{d^3p}{(2\pi)^3} 
\left[ E_p %+ \frac{\Sigma_++\Sigma_-}{2} 
+T \ln\left(1+{\rm e}^{-(E_p-\Sigma_+-\mu)/T} \right) 
+T \ln\left(1+{\rm e}^{-(E_p+\Sigma_-+\mu)/T} \right) 
\right]
 ~, 
\end{equation}
with the quasiparticle energy shift for quarks (antiquarks) due to mesonic correlations given by 
$\Sigma_\pm=\sum_{M=\pi,\sigma} {\rm tr}_D \left[ \Sigma_M\Lambda_\pm \gamma_0\right]/2$ and 
the positive (negative) energy projection operators $\Lambda_\pm=(1\pm \gamma_0)/2$.
The mesonic contribution to the thermodynamics is
\begin{equation}
\Omega_M = d_M \int \frac{d^3 k}{(2\pi)^3} \int\frac{d\omega}{2\pi}
\left\{
\omega + 2 T \ln \left[1 - {\rm e}^{-\omega/T} \right] 2\sin^2\delta_M(k,\omega) ~
\frac{\delta_M(k,\omega)}{d\omega}
\right\}~,
\end{equation}
where similar to the case of deuterons in nuclear matter the factor $2\sin^2\delta_M$ accounts for the 
fact that mesonic correlations in the continuum are partly already accounted for by the selfenergies
$\Sigma_M$ defining the improved selfconsistent quasiparticle picture.  
In the previous works of  Refs.~\cite{Hufner:1994ma,Zhuang:1994dw,Blaschke:2013zaa,Yamazaki:2012ux,Wergieluk} on this topic, however, the effect of the backreaction from mesonic correlations on the quark meanfield thermodynamics had been disregarded. 
Here we note from the $\Phi-$ derivable approach that for consistency the quark propagator in the quark meanfield thermodynamic potential shall contain effects from the selfenergy $\Sigma_M$ due to the coupling to the mesonic correlations as in the right panel of Fig.~\ref{fig:3}. 
This total quark selfenergy is the given by $\Sigma({\bf p},p_0)=\sigma_{\rm MF} + \Sigma_M({\bf p},p_0)$,
where for a local NJL model with scalar coupling constant $G_S$ the meanfield contribution is 
\begin{eqnarray}
\sigma_{\rm MF}= 2 N_f N_c G_S \int \frac{d^3 p}{(2\pi)^3} \frac{m}{E_p}[1-f_-(E_p)-f_+(E_p)] ~,
\end{eqnarray} 
and the contribution due to scalar/pseudoscalar mesons (corresponding to the diagram shown in the rightmost panel of Fig.~\ref{fig:3}) is given by \cite{Kitazawa:2014sga}
\begin{eqnarray}
\label{SigmaM}
\Sigma_M({\bf 0},p_0)= d_M\int \frac{d^4q}{(2\pi)^4}\pi \varrho_M({\bf q},q_0) 
\left\{\frac{(\gamma_0 +m/E_q)[1+g(q_0)-f_-(E_q)]}{q_0-p_0+E_q-\mu-i\eta}
+\frac{(\gamma_0 -m/E_q)[g(q_0)+f_+(E_q)]}{q_0-p_0-E_q-\mu-i\eta}
\right\}~,
\nonumber\\
\end{eqnarray} 
where $\varrho_M=(-1/\pi)\Im T_M({\bf q}, \omega+i\eta)$ is the meson spectral density and 
$E_q=\sqrt{q^2+m^2}$ is the quark dispersion law with the quark mass $m=m_0+\sigma_{\rm MF}$.
One can observe the similarity of this result (\ref{SigmaM}) with that for a Dirac fermion coupled to a 
pointlike scalar meson, as given in \cite{Blaizot:1991kh}.

%\subsection{Clusters in nuclear matter}

\subsection{Hadrons in quark matter}

\begin{figure}[!htb]
\includegraphics[width=0.19\textwidth]{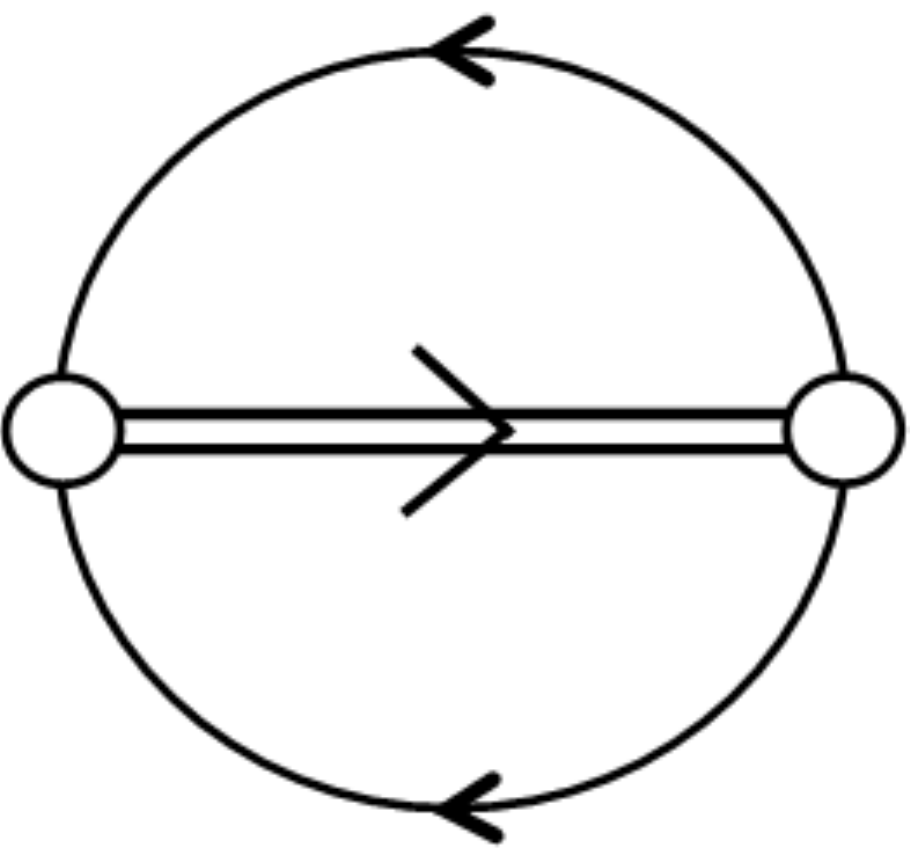} %\hfill
\includegraphics[width=0.19\textwidth]{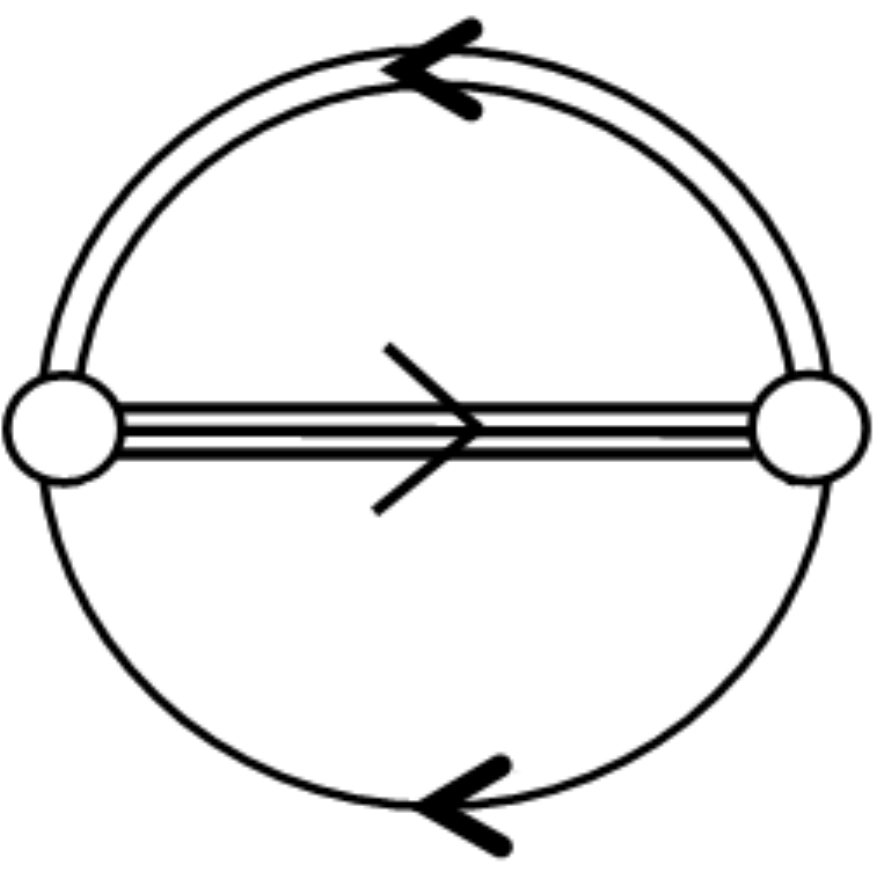}
%\\
\includegraphics[width=0.19\textwidth]{Fig5C}
\includegraphics[width=0.19\textwidth]{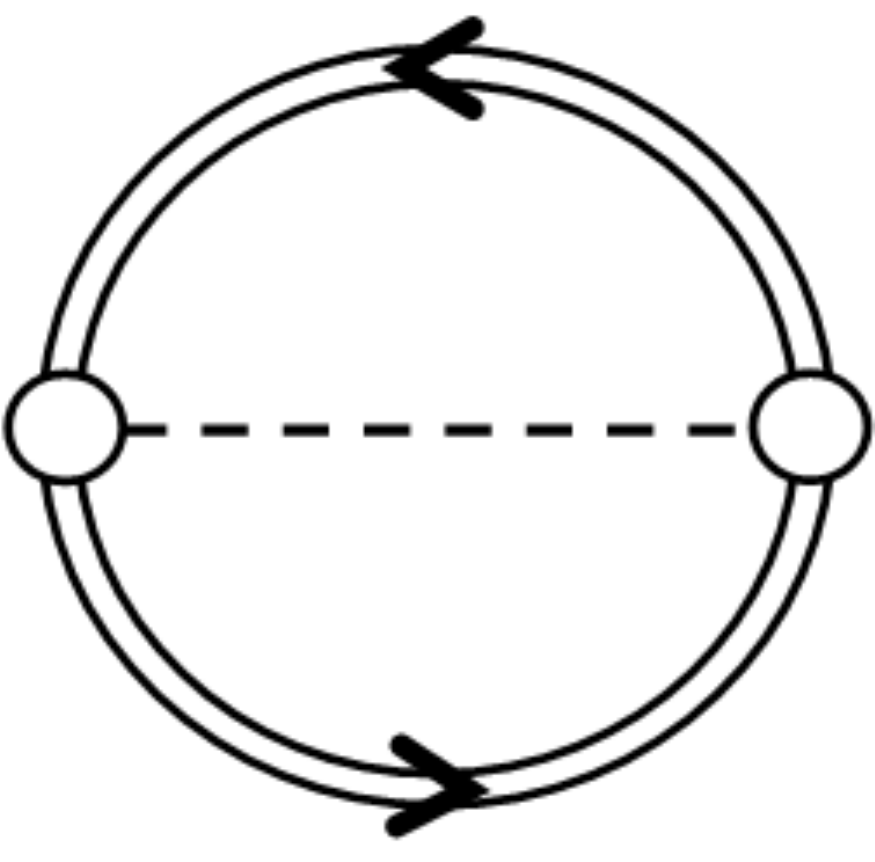}
\includegraphics[width=0.19\textwidth]{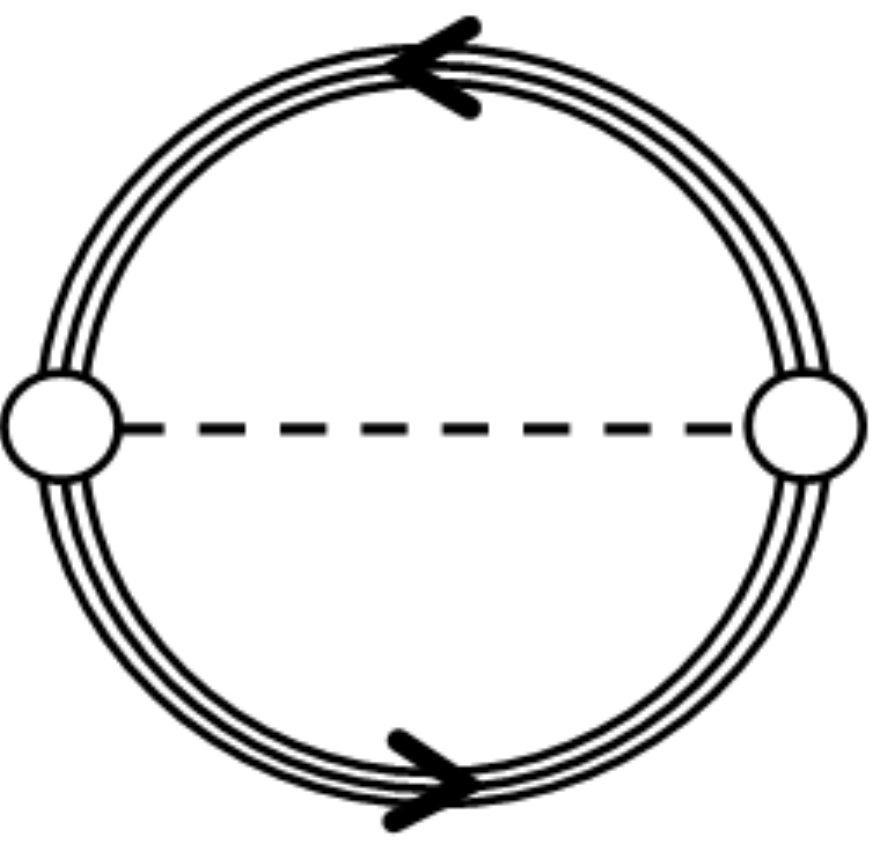}
\caption{The contributions to the $\Phi-$ functional for the quark-meson-diquark-baryon system.} 
\label{fig:4}
\end{figure}

\begin{figure}[!htb]
\includegraphics[width=0.19\textwidth]{Fig6I} %\hfill
\includegraphics[width=0.19\textwidth]{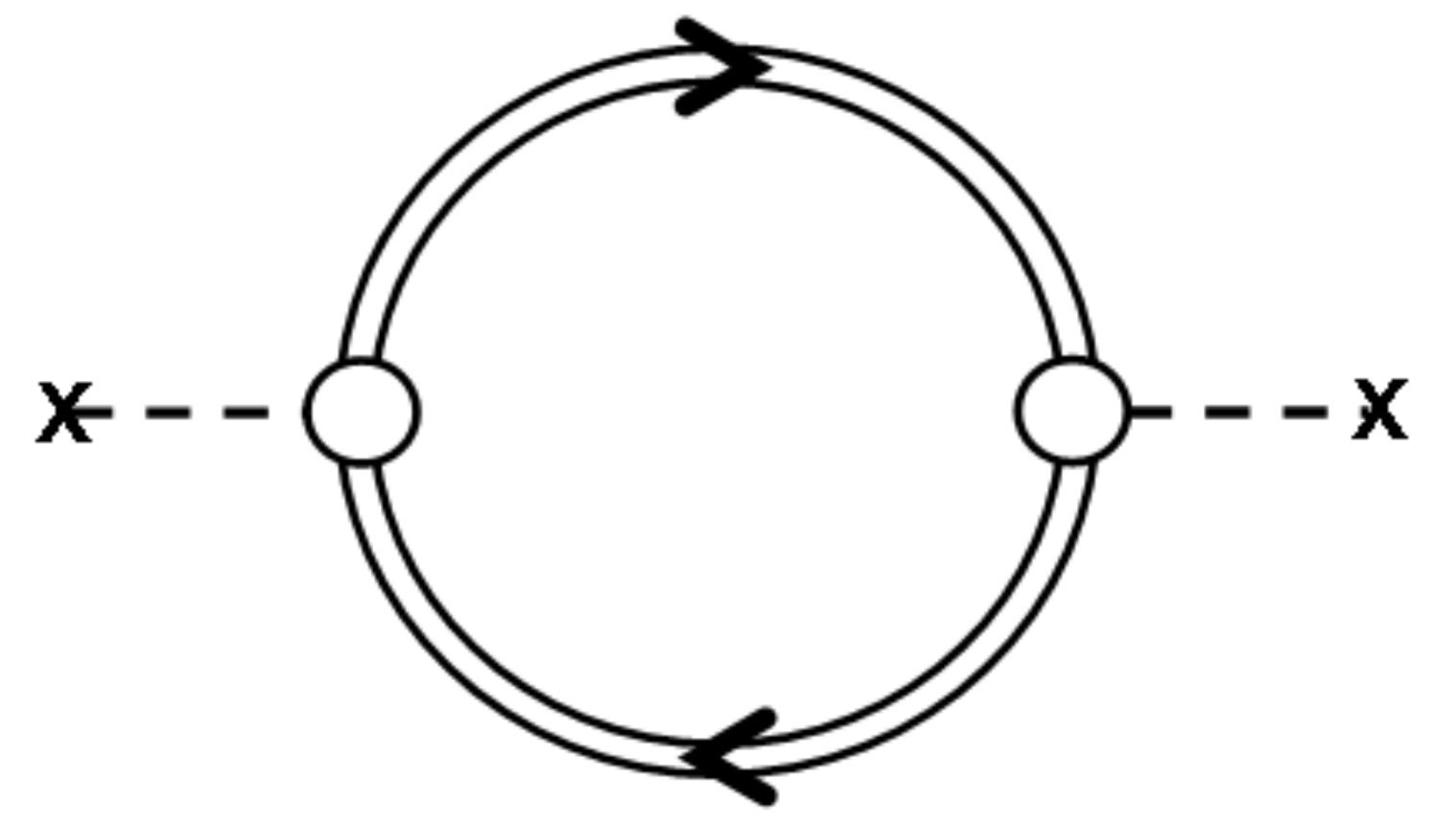}
\includegraphics[width=0.19\textwidth]{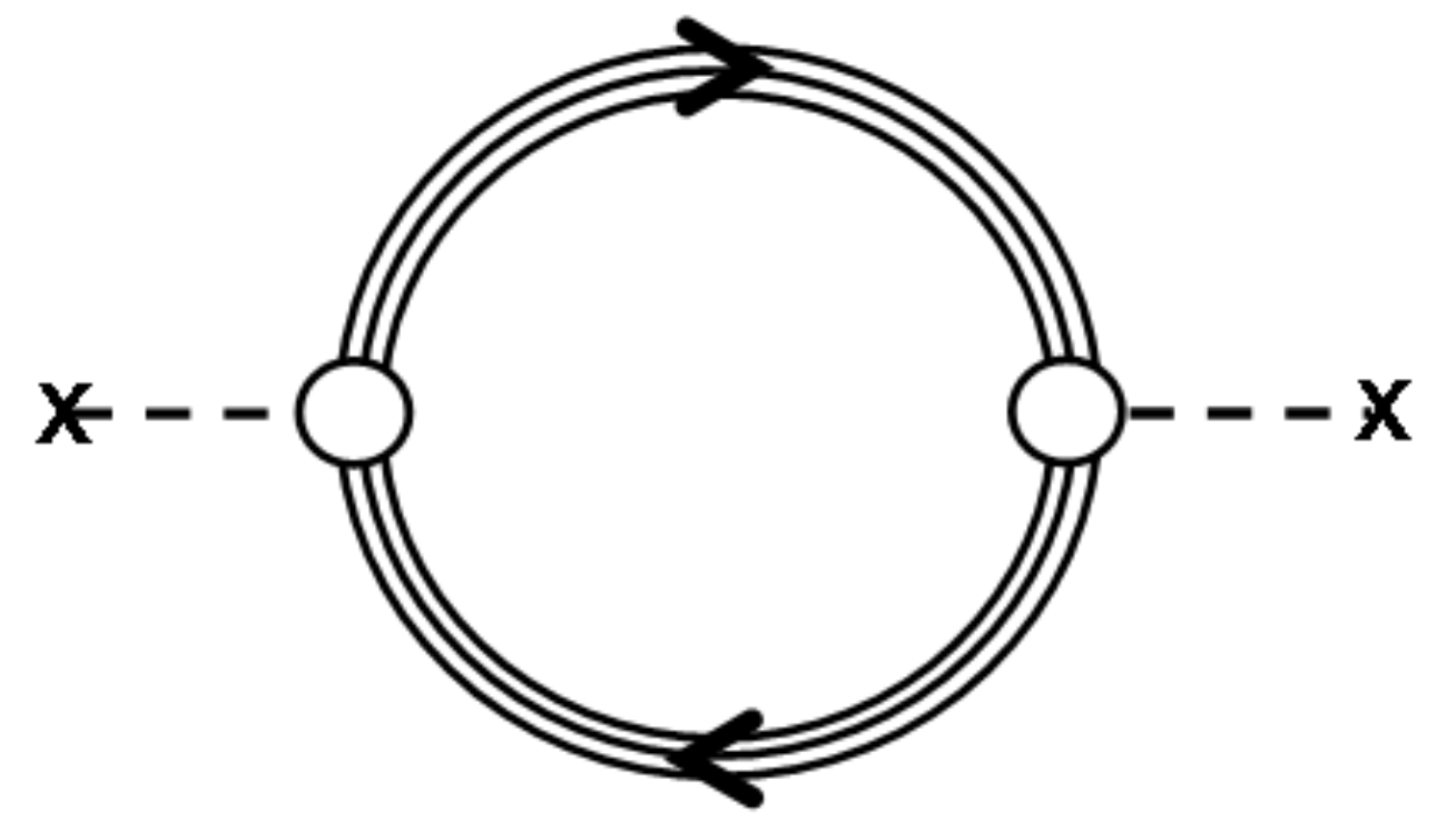}
\\
\includegraphics[width=0.19\textwidth]{Fig6D} %\hfill
\includegraphics[width=0.19\textwidth]{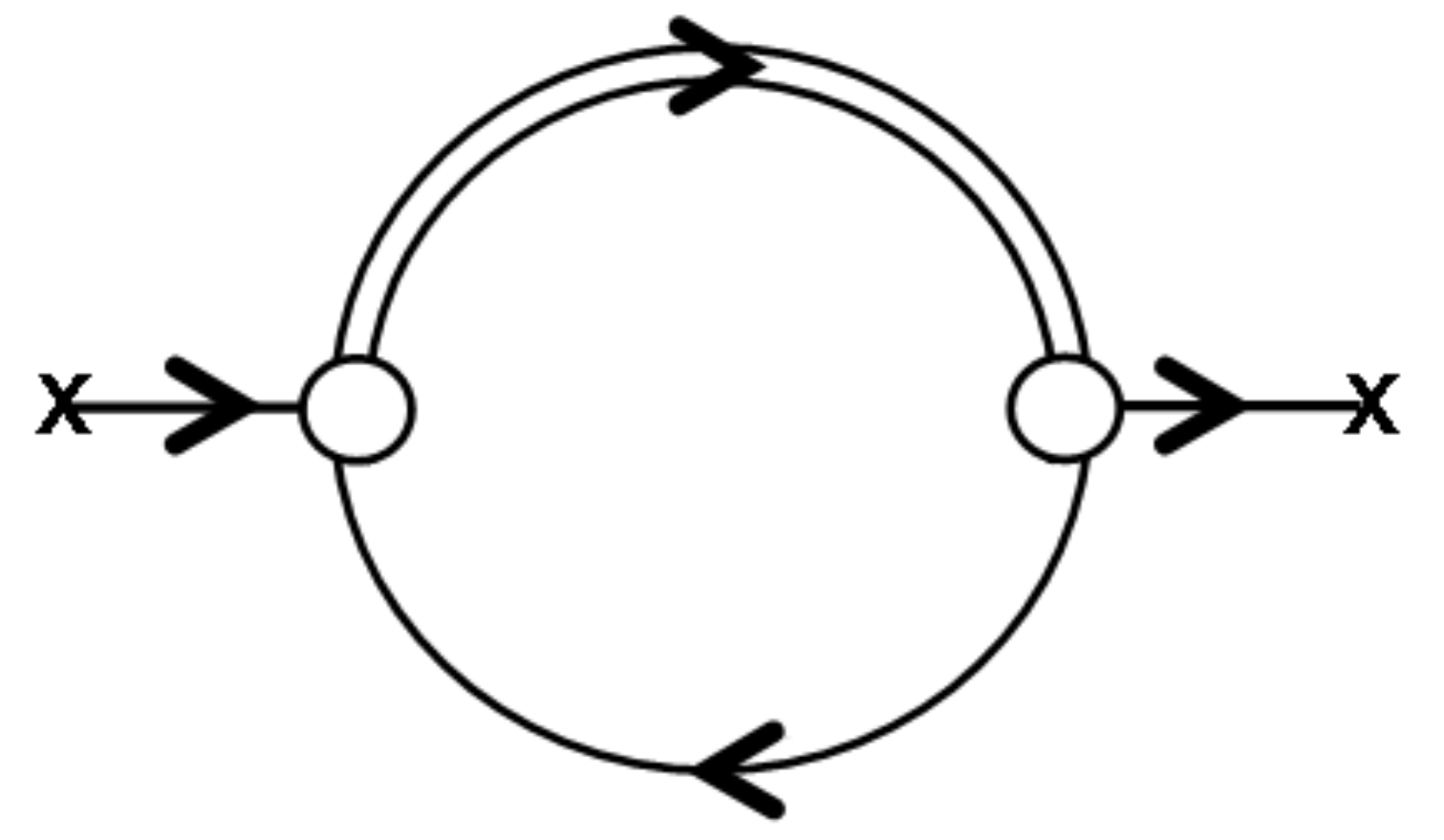}
\includegraphics[width=0.19\textwidth]{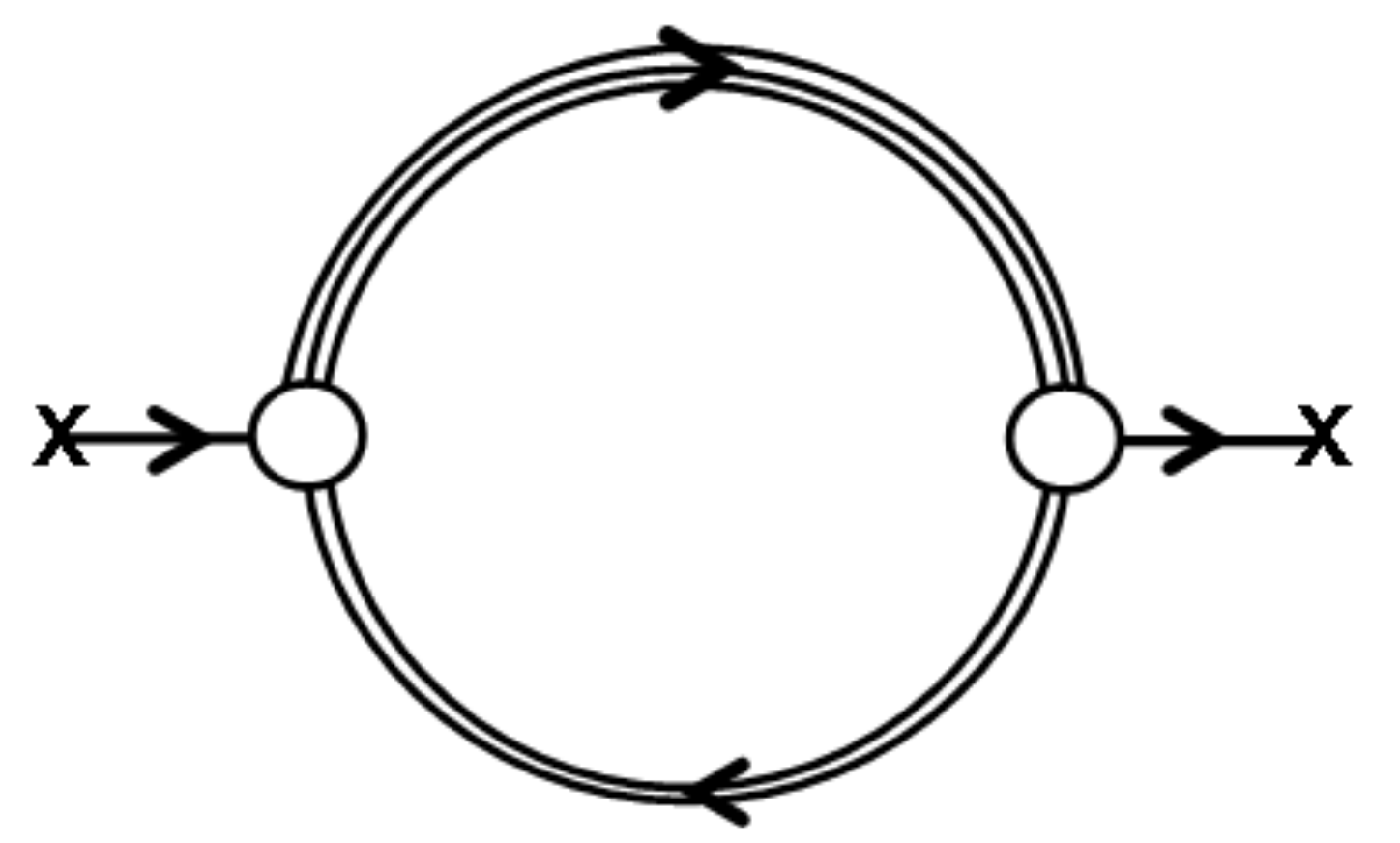}
\\
\includegraphics[width=0.19\textwidth]{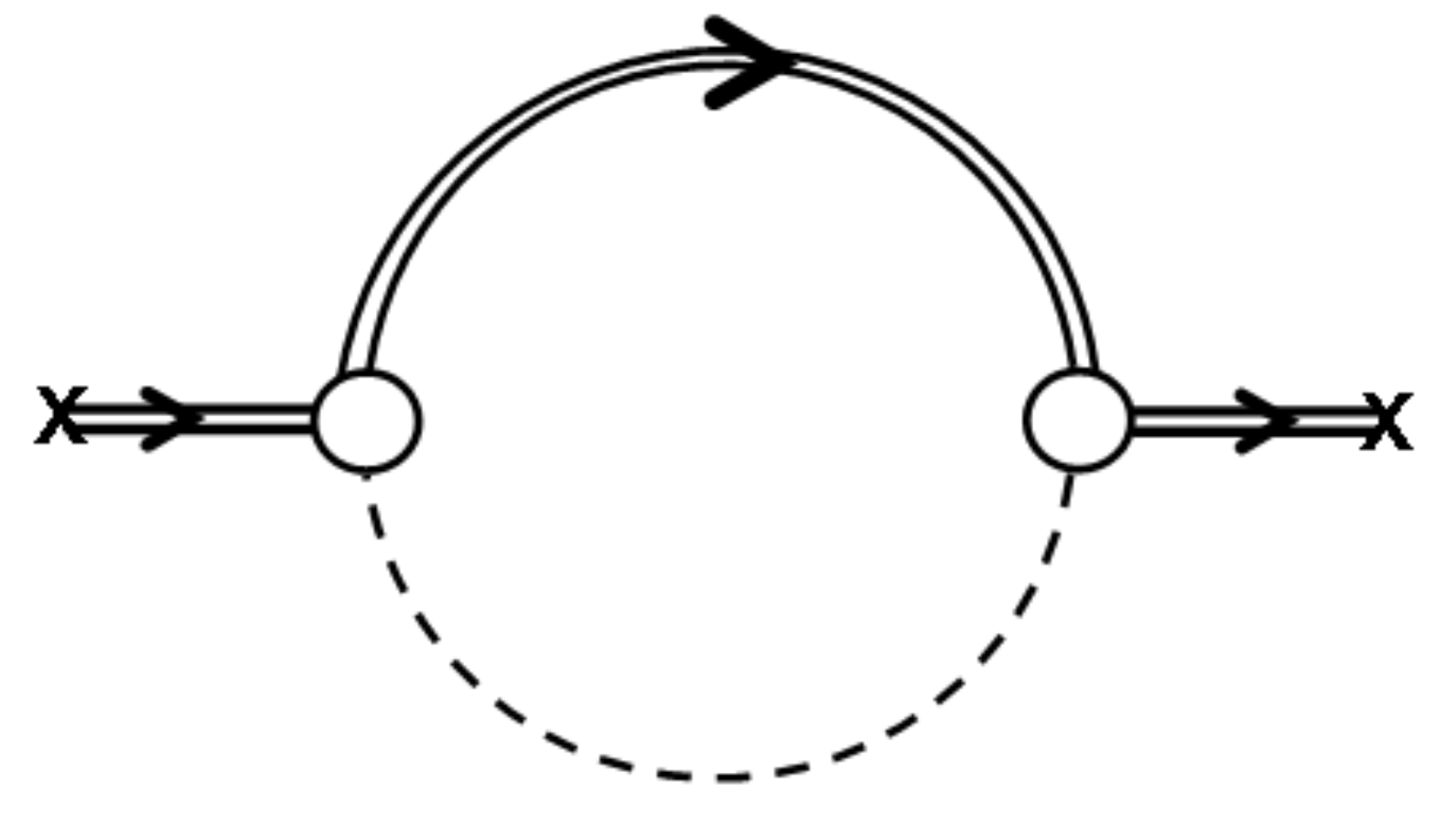}
\includegraphics[width=0.19\textwidth]{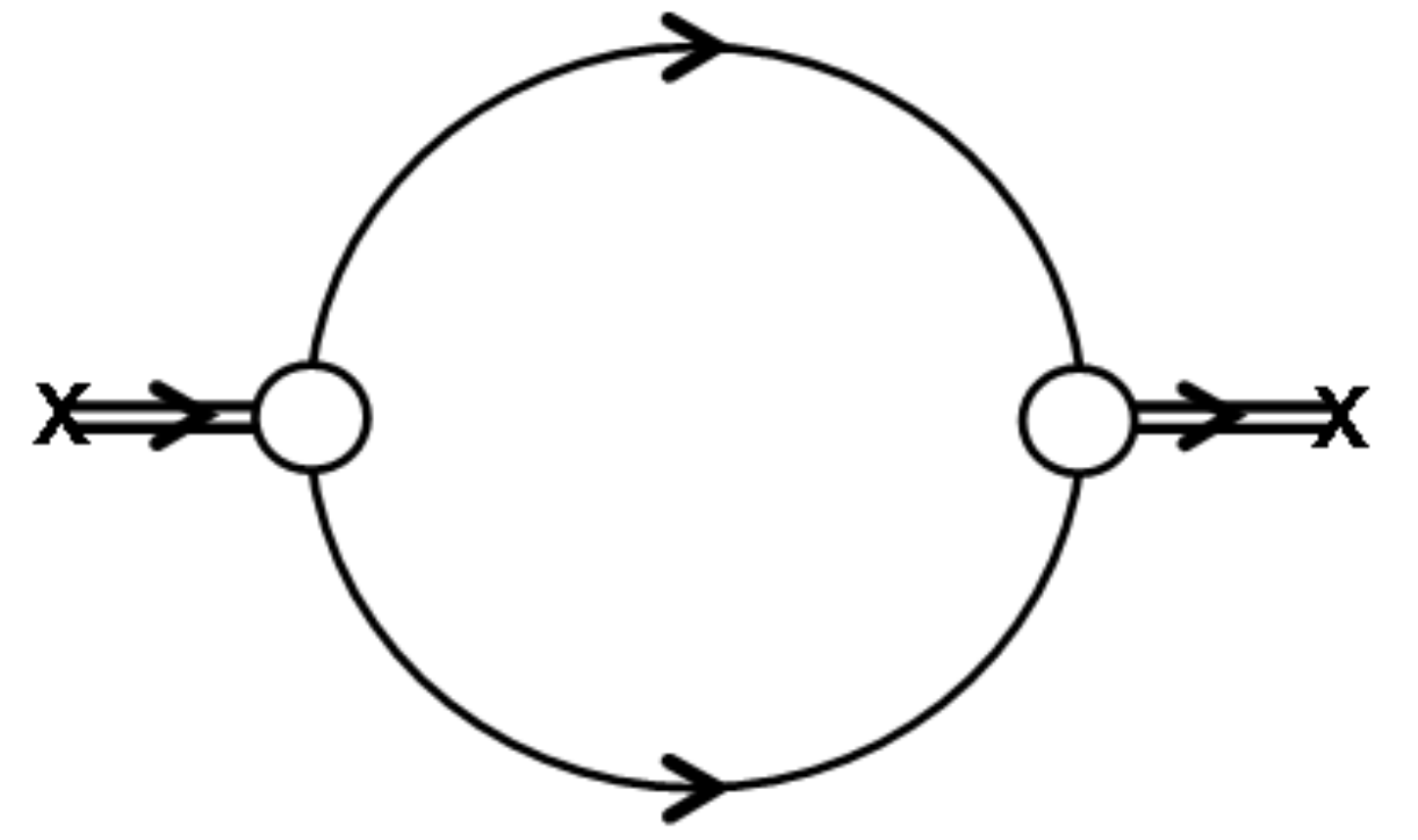}
\includegraphics[width=0.19\textwidth]{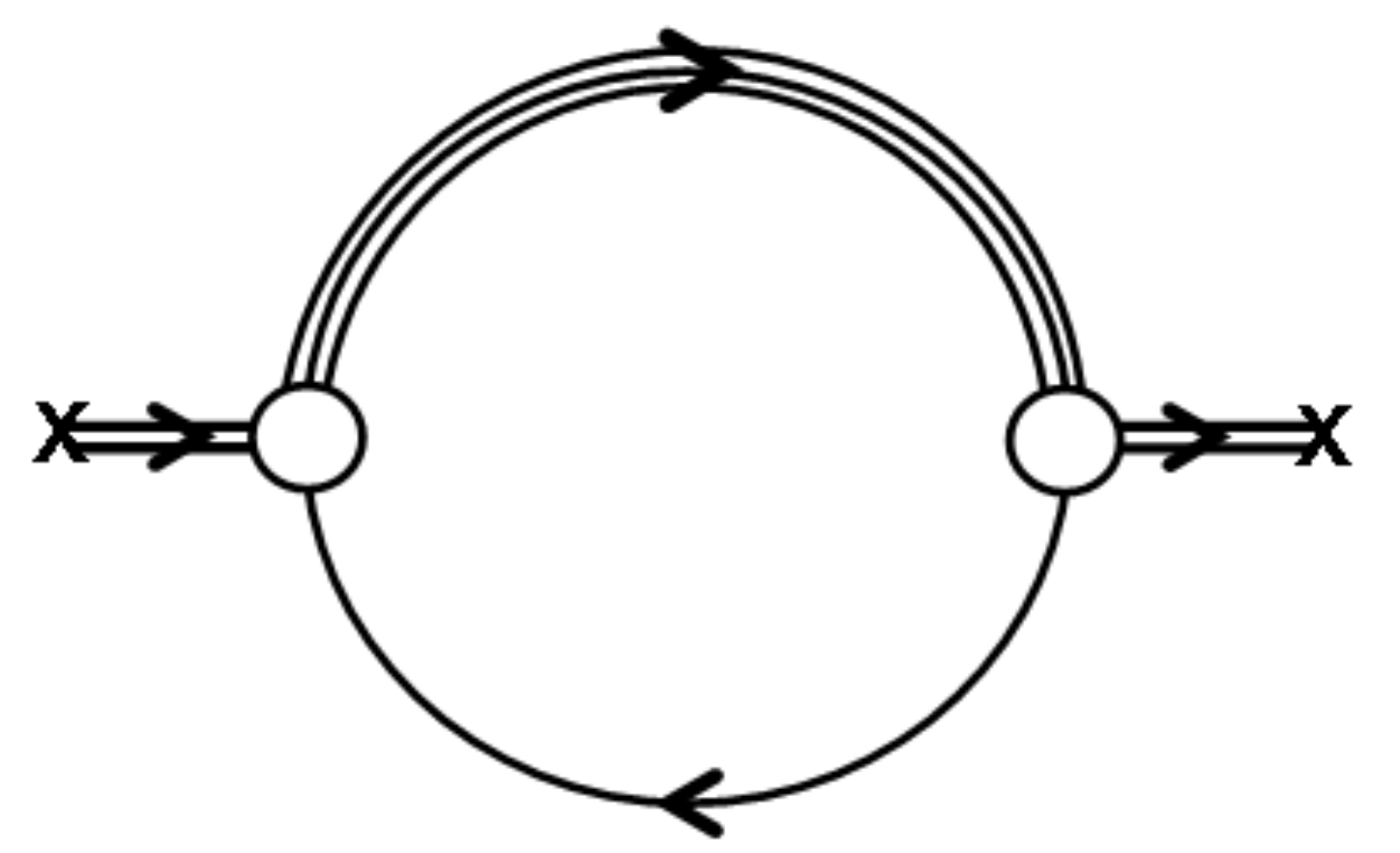}
\\
\includegraphics[width=0.19\textwidth]{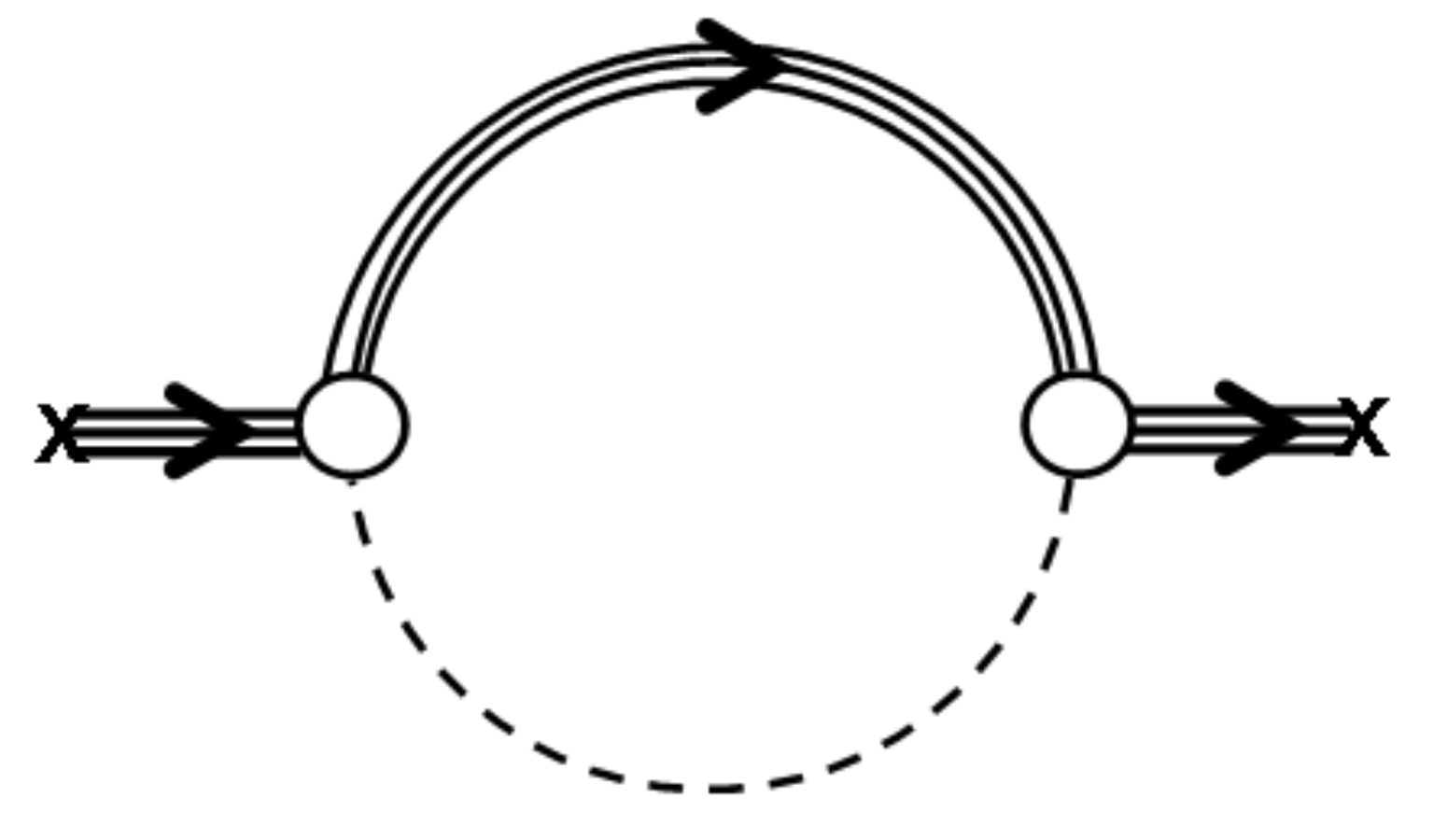}
\includegraphics[width=0.19\textwidth]{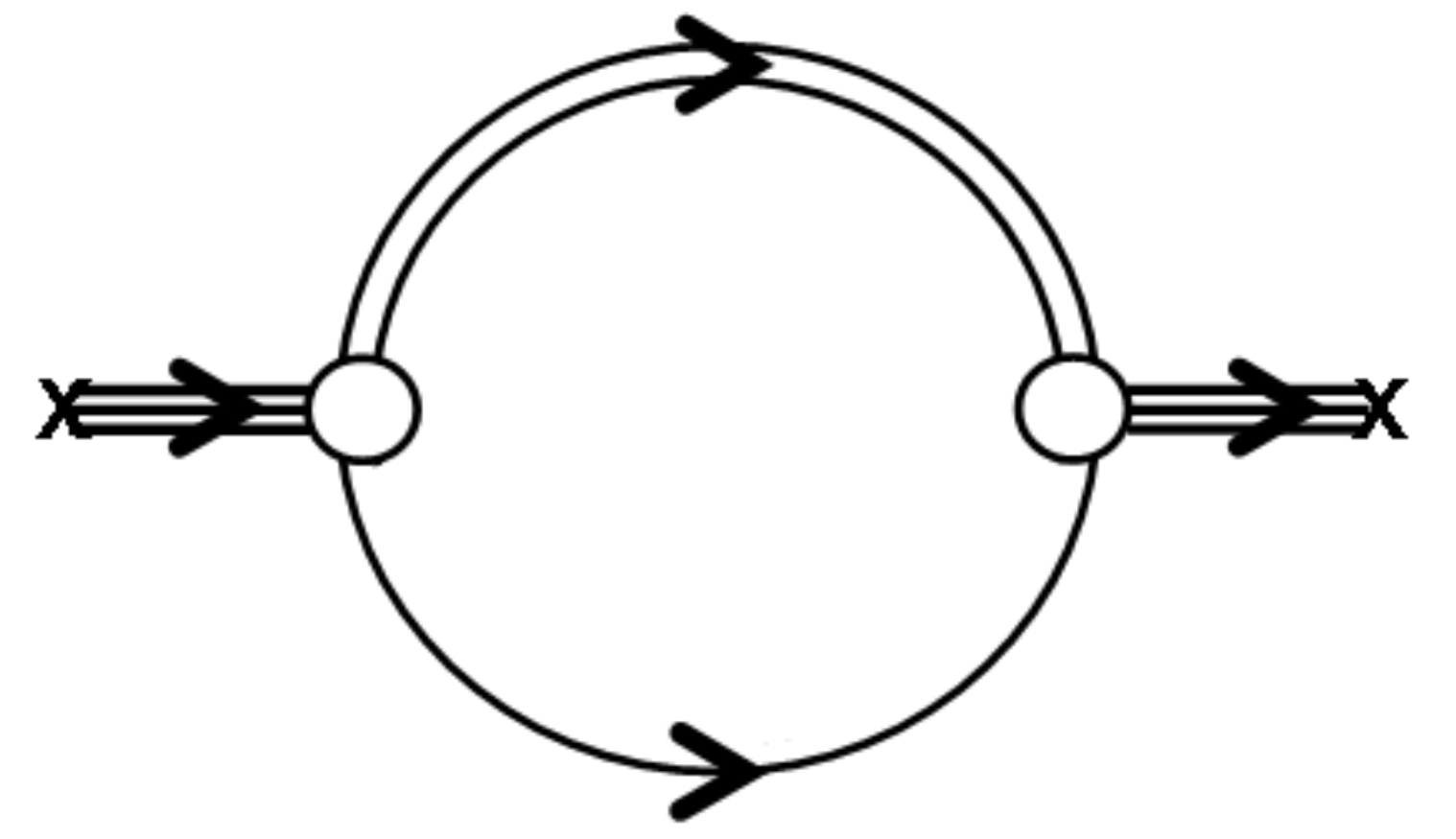}
\caption{The selfenergy contributions for the Greens functions of the quark-meson-diquark-baryon system,
defined by the $\Phi-$ functional contributions shown in Fig.~4. 
From top to down the four rows of diagrams show the selfenergies for the full propagators of mesons, quarks, diquarks and baryons, respectively.
\label{fig:5}}
\end{figure}

Finally, we would like to sketch how the $\Phi-$ derivable approach can be employed to define a cluster virial expansion for quark-hadron matter consisting of quarks (Q), mesons (M), diquarks (D) and baryons (B). The thermodynamical potential for this system obtains the form very similar to the case of clustered nuclear matter, i.e.
\begin{eqnarray}
\Omega &=& \sum_{i=Q,M,D,B} (-1)^{c_i} \left[{\rm Tr} \ln \left(-G_i^{-1}\right) 
+ {\rm Tr} \left(\Sigma_i~G_i \right) \right]+\Phi\left[G_Q,G_M,G_D,G_B\right]~,
\end{eqnarray}
where $c_i=0$ ($c_i=1$) for bosonic (fermionic) states and the minimal set of two-particle irreducible diagrams defining the   $\Phi-$ functional is given in Fig.~\ref{fig:4}.
From this $\Phi-$ functional follow the selfenergies defining the full Greens functions of the system by functional derivation
\begin{equation}
\Sigma_i = \frac{\delta~ \Phi\left[G_Q,G_M,G_D,G_B\right]}{\delta~G_i }~.
\end{equation}
The resulting Feynman diagrams for the selfenergy contributions are given in Fig.~\ref{fig:5}.
Note that it is immediately plain from this formulation that in the situation of confinement, when the propagators belonging to colored excitations (quarks and diquarks) and thus to states that could not be populated, the system simplifies considerably. When all closed loop diagrams containing quarks and diquarks are neglected, this system reduces to a meson-baryon system as, e.g., described in selfconsistent relativistic meanfield theories of nuclear matter.
On the other hand, in the case of deconfined quarks when also chiral symmetry is restored, the meson and baryon states become unbound (Mott effect) and their contribution to the thermodynamics as captured in the corresponding phase shift functions is gradually vanishing at high temperatures and chemical potentials with just chiral quark matter remaining asymptotically.

\section{Conclusion}
In this contribution the $\Phi-$ derivable approach is generalized to describe the formation of bound states (clusters) of different size in many particle systems with strong interactions.  
A generic form of $\Phi-$ functionals is presented that turns out to be fully equivalent to a selfconsistent cluster virial expansion up to the second virial coefficient for interactions among the clusters.
As examples are considered: nuclei in nuclear matter and hadrons in quark matter, with particular attention to the case of the deuterons in nuclear matter and mesons in quark matter. 
Generalized Beth-Uhlenbeck equations of state are derived, where the quasiparticle virial expansion is
extended to include arbitrary clusters. 
The approach is applicable to nonrelativistic potential models of nuclear matter as well as to relativistic
field theoretic models of quark matter. 
It is particularly suited for a description of cluster formation and dissociation in hot, dense matter.

\subsection*{Acknowledgements}
Many ideas discussed in this contribution have emerged from a long standing collaboration with 
H. Grigorian, T. Kl\"ahn, G. R\"opke, S. Typel and H. Wolter who are gratefully acknowledged for 
insightful discussions. I thank N. T. Gevorgyan for excellent assistance in preparing this manuscript.
This work was supported in part  by the Polish National Science Center (NCN) under grant number
UMO-2011/02/A/ST2/00306.

%Polish Ministry of Science and Higher Education under grant No. 1009/S/IFT/14.


\begin{thebibliography}{99}

\bibitem{KB}
G. Baym, L. P. Kadanoff, Phys. Rev. {\bf 124}, 287 (1961); 
G. Baym, Phys. Rev. {\bf 127}, 1391 (1962).

\bibitem{David2}
  R. Dashen, S. Ma, W. J. Bernstein, 
  Phys. Rev. {\bf 187}, 345 (1969).

\bibitem{David3}
  J.M. Cornwall, R. Jackiw, E. Tomboulis,
  Phys. Rev. D {\bf 10}, 2428 (1974).

\bibitem{David4}
 G.M. Carneiro, C.J. Pethick, Phys. Rev. D {\bf 11}, 1106 (1975).


\bibitem{David5}
T.D. Lee, M. Margulies, Phys. Rev. D {\bf 11}, 1591 (1975).

\bibitem{David6}
E.M. Nyman, M. Rho, Nucl. Phys. A {\bf 268}, 408 (1976).

\bibitem{David1}
  W. Weinhold, B. Friman, W. N\"{o}renberg, 
  Phys. Lett. B {\bf 433}, 236 (1998).
%  in particular Ref.\ [7]-[9].

\bibitem{RMS}
  G.\ R\"opke, M.\ Schmidt, L.\ M\"unchow, and H.\ Schulz,
  Nucl. Phys. A {\bf 379}, 536 (1982);
  Nucl. Phys. A {\bf 399}, 587 (1983);
  Phys. Lett. {\bf B 110}, 21 (1982).

%\cite{Zimmermann:1985ji}
\bibitem{Zimmermann:1985ji}
  R.~Zimmermann, H.~Stolz,
 % ``The Mass Action Law in Two-Component Fermi Systems Revisited Excitons and 
  % Electron-Hole Pairs,''
  physica\ status\ solidi\ (b) {\bf 131}, 151 (1985).

\bibitem{SRS}
  M.~Schmidt, G.~R\"opke, and H.~Schulz,
  Ann. Phys. {\bf 202}, 57 (1990).

%\cite{Beth:1936zz}
\bibitem{Beth:1936zz}
  G.~Uhlenbeck, E.~Beth, 
%	``The quantum theory of the non-ideal gas. 
%	I. Deviations from the classical theory,''
  Physica {\bf 3}, 729 (1936).

%\cite{Beth:1937zz}
\bibitem{Beth:1937zz}
  E.~Beth, G.~Uhlenbeck,
%	``The quantum theory of the non-ideal gas. 
%	II. Behaviour at low temperatures,''
  Physica {\bf 4}, 915 (1937).

%\cite{Typel:2009sy}
\bibitem{Typel:2009sy} 
  S.~Typel, G.~R\"opke, T.~Kl\"ahn, D.~Blaschke and H.~H.~Wolter,
  %``Composition and thermodynamics of nuclear matter with light clusters,''
  Phys.\ Rev.\ C {\bf 81}, 015803 (2010).
%  [arXiv:0908.2344 [nucl-th]].
  %%CITATION = ARXIV:0908.2344;%%

%\cite{Ropke:2014fia}
\bibitem{Ropke:2014fia} 
  G.~R\"opke,
  %``Nuclear matter equation of state including few-nucleon correlations $(A\leq 4)$,''
  arXiv:1411.4593 [nucl-th].
  %%CITATION = ARXIV:1411.4593;%%

%\cite{Ropke:2012qv}
\bibitem{Ropke:2012qv} 
  G.~R\"opke, N.-U.~Bastian, D.~Blaschke, T.~Kl\"ahn, S.~Typel and H.~H.~Wolter,
  %``Cluster virial expansion for nuclear matter within a quasiparticle statistical approach,''
  Nucl.\ Phys.\ A {\bf 897}, 70 (2013).
%  [arXiv:1209.0212 [nucl-th]].
  %%CITATION = ARXIV:1209.0212;%%

%\cite{Blaizot:1991kh}
\bibitem{Blaizot:1991kh} 
  J.~P.~Blaizot,
  %``Quantum fields at finite temperature and density,''
  J.\ Korean Phys.\ Soc.\  {\bf 25}, S65 (1992).
  %%CITATION = JKPSD,25,S65;%%

%\cite{Kitazawa:2014sga}
\bibitem{Kitazawa:2014sga} 
  M.~Kitazawa, T.~Kunihiro and Y.~Nemoto,
  %``Emergence of soft quark excitations by the coupling with a soft mode of the QCD critical point,''
  Phys.\ Rev.\ D {\bf 90}, 116008 (2014).
%  [arXiv:1409.3733 [hep-ph]].
  %%CITATION = ARXIV:1409.3733;%%

%\cite{Hufner:1994ma}
\bibitem{Hufner:1994ma} 
  J.~H\"ufner, S.~P.~Klevansky, P.~Zhuang and H.~Voss,
  %``Thermodynamics of a quark plasma beyond the mean field: A generalized Beth-Uhlenbeck approach,''
  Annals Phys.\  {\bf 234}, 225 (1994).
  %%CITATION = APNYA,234,225;%%

%\cite{Zhuang:1994dw}
\bibitem{Zhuang:1994dw} 
  P.~Zhuang, J.~H\"ufner and S.~P.~Klevansky,
  %``Thermodynamics of a quark - meson plasma in the Nambu-Jona-Lasinio model,''
  Nucl.\ Phys.\ A {\bf 576}, 525 (1994).
  %%CITATION = NUPHA,A576,525;%%

%\cite{Blaschke:2013zaa}
\bibitem{Blaschke:2013zaa} 
  D.~Blaschke, D.~Zablocki, M.~Buballa, A.~Dubinin and G.~R\"opke,
  %``Generalized Beth--Uhlenbeck approach to mesons and diquarks in hot, dense quark matter,''
  Annals Phys.\  {\bf 348}, 228 (2014).
%  [arXiv:1305.3907 [hep-ph]].
  %%CITATION = ARXIV:1305.3907;%%

%\cite{Yamazaki:2012ux}
\bibitem{Yamazaki:2012ux} 
  K.~Yamazaki and T.~Matsui,
  %``Quark-Hadron Phase Transition in the PNJL model for interacting quarks,''
  Nucl.\ Phys.\ A {\bf 913}, 19 (2013).
%  [arXiv:1212.6165 [hep-ph]].
  %%CITATION = ARXIV:1212.6165;%%

%\cite{Wergieluk:2012gd}
\bibitem{Wergieluk:2012gd} 
  A.~Wergieluk, D.~Blaschke, Y.~L.~Kalinovsky and A.~Friesen,
  %``Pion dissociation and Levinson`s theorem in hot PNJL quark matter,''
  Phys.\ Part.\ Nucl.\ Lett.\  {\bf 10}, 660 (2013).
  %[arXiv:1212.5245 [nucl-th]].
  %%CITATION = ARXIV:1212.5245;%%

\end{thebibliography}
\end{document}